\documentclass[10pt, conference, letterpaper]{IEEEtran}
\IEEEoverridecommandlockouts

\usepackage{amsmath,amssymb,amsfonts}
\usepackage{algorithmic}
\usepackage{algorithm}
\usepackage{array}
\usepackage{textcomp}
\usepackage{stfloats}
\usepackage{url}
\usepackage{verbatim}
\usepackage{graphicx}
\usepackage{cite}
\usepackage{subcaption}
\usepackage{xcolor}
\usepackage{float}
\usepackage{enumitem}
\usepackage{placeins}

\def\BibTeX{{\rm B\kern-.05em{\sc i\kern-.025em b}\kern-.08em
    T\kern-.1667em\lower.7ex\hbox{E}\kern-.125emX}}
\begin{document}

\title{Criticality-Based Dynamic Topology Optimization for Enhancing Aerial-Marine Swarm Resilience}

\author{
\IEEEauthorblockN{
Ruiyang Huang\IEEEauthorrefmark{2}\IEEEauthorrefmark{4},
Haocheng Wang\IEEEauthorrefmark{2}\IEEEauthorrefmark{4},
Yixuan Shen\IEEEauthorrefmark{2},
Ning Gao\IEEEauthorrefmark{2}\IEEEauthorrefmark{1},
Qiang Ni\IEEEauthorrefmark{3},
Shi Jin\IEEEauthorrefmark{2},
and Yifan Wu\IEEEauthorrefmark{5}}
\IEEEauthorblockA{\IEEEauthorrefmark{2}Southeast University. Email: \{ryhuang\_572, 213232710, 213232137, ninggao, jinshi\}@seu.edu.cn}
\IEEEauthorblockA{\IEEEauthorrefmark{3}Lancaster University. Email: q.ni@lancaster.ac.uk}
\IEEEauthorblockA{\IEEEauthorrefmark{5}Peking University. Email: yifanwu@pku.edu.cn}
\thanks{§\textit{ Ruiyang Huang and Haocheng Wang are co-first authors.}

* \textit{Corresponding author: Ning Gao.}}
}

\maketitle

\begin{abstract}
Heterogeneous marine-aerial swarm networks encounter substantial difficulties due to targeted communication disruptions and structural weaknesses in adversarial environments. This paper proposes a two-step framework to strengthen the network's resilience. Specifically, our framework combines the node prioritization based on criticality with multi-objective topology optimization. First, we design a three-layer architecture to represent structural, communication, and task dependencies of the swarm networks. Then, we introduce the SurBi-Ranking method, which utilizes graph convolutional networks, to dynamically evaluate and rank the criticality of nodes and edges in real time. Next, we apply the NSGA-III algorithm to optimize the network topology, aiming to balance communication efficiency, global connectivity, and mission success rate. Experiments demonstrate that compared to traditional methods like K-Shell, our SurBi-Ranking method identifies critical nodes and edges with greater accuracy, as deliberate attacks on these components cause more significant connectivity degradation. Furthermore, our optimization approach, when prioritizing SurBi-Ranked critical components under attack, reduces the natural connectivity degradation by around 30\%, achieves higher mission success rates, and incurs lower communication reconfiguration costs, ensuring sustained connectivity and mission effectiveness across multi-phase operations.
\end{abstract}


\section{Introduction}
\subsection{Background and Motivation}
Heterogeneous marine-aerial swarm networks, integrating unmanned aerial vehicles (UAVs) and unmanned surface vehicles (USVs), have emerged as critical platforms for applications such as search-and-rescue (SAR), environmental surveillance, and reconnaissance. These networks leverage the complementary capabilities of UAVs and USVs to achieve superior coverage, flexibility, and efficiency in challenging marine and aerial environments \cite{liu_adaptive_2023}, \cite{tang_mission-oriented_2024}. Their ability to operate collaboratively in remote or hazardous settings makes them indispensable for mission-critical tasks.

However, these networks face significant challenges in adversarial environments, where physical layer security is a primary concern. Targeted attacks, such as signal jamming and physical disruptions, exploit vulnerabilities in network topology, compromising communication connectivity and mission integrity \cite{li_baseline-resilience_2022}, \cite{zhou_resilient_2022}. Such threats can sever critical communication links or disable key nodes, leading to cascading failures that undermine the swarm’s operational effectiveness. Traditional resilience approaches, often relying on static or single-layer models, fail to address the complex interdependencies across structural, communication, and task layers \cite{kabashkin_resilience_2024}, \cite{beck_guidance_2022}, leaving swarms vulnerable to sophisticated attacks \cite{su_heterogeneous_2024}, \cite{phadke_increasing_2024}.
\begin{figure}[t]
    \centering
    \includegraphics[width=\columnwidth]{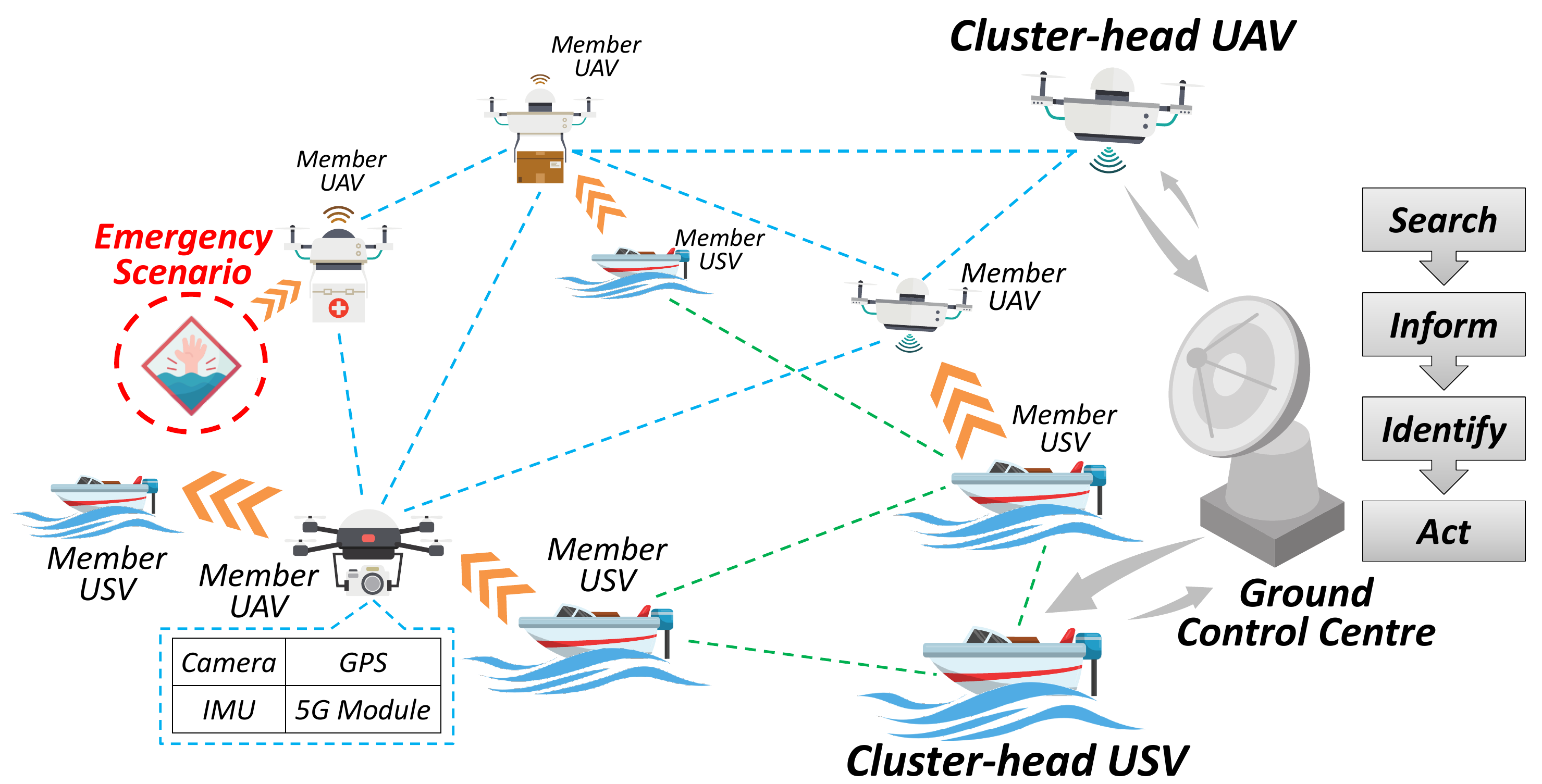}
    \caption{Heterogeneous marine–aerial swarm for SAR missions.}
    \label{fig1}
\end{figure}

To this end, enhancing the robustness, security, and attack resilience of heterogeneous marine-aerial swarm networks through advanced topology optimization requires addressing the following fundamental questions:

\begin{enumerate}[label= \textbf{\textit{P\arabic*)}} ]
    \item \textbf{\textit{How can cross-layer dependencies be effectively modeled to mitigate adversarial threats?}} Current models often overlook the interplay between structural formations, communication reliability, and task dependencies \cite{chen_cross-layer_2023}, \cite{sahoo_multilayer_2021}, \cite{li_applying_2024}, limiting their ability to predict and prevent cascading failures in contested environments.

    \item \textbf{\textit{How can critical nodes and links be accurately identified across multiple layers?}} Traditional centrality metrics like K-Shell and single Birnbaum importance fail to capture dynamic, cross-layer node importance and neighboring influences, limiting effective protection against targeted attacks in swarm networks \cite{min_evaluation_2022}, \cite{niu_identification_2020}, \cite{wang_topology_2020}.

    \item \textbf{\textit{How can network topology be optimized to balance robustness, reliability, and mission success?}} Achieving a topology that simultaneously ensures structural connectivity, communication security, and mission effectiveness under resource constraints and adversarial threats remains a significant challenge \cite{kim_bottlenet_2021}.
\end{enumerate}

Based on the three problems above, our research is motivated by the urgent need to enhance the attack resilience of heterogeneous marine-aerial swarm networks and ensure the effective execution of multi-phase missions in contested environments. Addressing the challenges of modeling cross-layer dependencies, accurately identifying critical nodes and links, and optimizing network topology is essential to strengthen swarm connectivity, mitigate adversarial threats, and maintain mission continuity under dynamic and hostile conditions.

\subsection{Solution and Contributions}
In this paper, we propose a comprehensive framework aimed at improving network robustness via criticality analysis and topology optimization. Specifically, our approach integrates a three-layer network model, an advanced adversarial model, a criticality assessment method, and multi-objective topology optimization. The main contributions of this paper are summarized as follows:
\begin{itemize}
    \item \textbf{\textit{Three-layer network modeling and adversarial modeling:}} We propose a three-layer network model that captures the structural, communication, and task layers of marine-aerial swarms. The structural layer encodes physical formations, the communication layer models probabilistic link failures based on inter-node distances, and the task layer represents mission dependencies using percolation-based thresholds. Complementing this, we design a sophisticated adversarial model simulating a powerful attacker with situational awareness, capable of launching targeted node and edge disruptions, including signal jamming and physical attacks, to induce cascading failures and disrupt mission continuity.
    \item \textbf{\textit{SurBi-Ranking method for criticality assessment:}} We propose a SurBi-Ranking method, which combines graph convolutional networks (GCNs) with complex network theory to evaluate the criticality of nodes and edges. By integrating Birnbaum importance and surrounding node influence, this method accurately identifies critical components whose failure significantly impacts connectivity and mission success, enabling prioritized protection strategies.
    \item \textbf{\textit{Multi-objective topology optimization:}} Leveraging the NSGA-III algorithm, we optimize swarm network topologies to balance structural robustness, communication reliability, and mission success probability. Our approach obtains the Pareto-optimal topology that minimizes reconfiguration costs while enhancing resilience against adversarial attacks, ensuring sustained operational effectiveness across diverse mission scenarios.
\end{itemize}

To perform the experiments, our code is available at https://anonymous.4open.science/r/TopoOptimEC24/ .
\subsection{Related Works}
We review recent advancements in system modeling, criticality assessment, and topology optimization, identifying gaps addressed by our proposed framework.

\subsubsection{\textbf{\textit{Threats and Resilience in Unmanned Swarm Networks}}}
Recent studies emphasize multi-layer modeling to enhance resilience in UAV-USV swarms. Liu et al. \cite{liu_leveraging_2021} proposed double-layer coupled models and mission-oriented metrics to capture attack impacts and support recovery, validated in realistic scenarios. Guo et al.  and Zhang et al. \cite{zhang_uav-enabled_2024} [9], [11] introduced spatio-temporal dynamics and cascading failure models, improving resilience assessment. However, these approaches often focus on static or dual-layer models, neglecting comprehensive inter-layer dependencies and sophisticated adversarial strategies, limiting their effectiveness in contested environments.
Recent studies emphasize multi-layer modeling to enhance resilience in UAV-USV swarms. Liu et al. \cite{liu_leveraging_2021} proposed double-layer coupled models and mission-oriented metrics to capture attack impacts and support recovery, validated in realistic scenarios. Guo et al. \cite{guo_resilience_2024} introduced spatio-temporal dynamics and cascading failure models, improving resilience assessment. However, these approaches mainly focus on static or dual-layer models, neglecting comprehensive inter-layer dependencies and sophisticated adversarial strategies, limiting their effectiveness in contested environments.

\subsubsection{\textbf{\textit{Criticality Assessment Methods}}}
Criticality assessment is key to identifying vulnerable nodes and links. Methods leveraging topological indicators including node degree and clustering coefficient, and dynamic evolution models have been developed to evaluate component importance under attack scenarios \cite{zhang_research_2021}, \cite{li_baseline-resilience_2022}. Dui et al. \cite{dui_importance-based_2024} proposed a framework integrating preventive, robustness, and recoverability metrics for UAV and USV swarms, enabling targeted protection. Yet, these methods often rely on traditional metrics, lacking the ability to capture dynamic, multi-layer criticality, which is essential for prioritizing defenses in complex swarms.

\subsubsection{\textbf{\textit{Topology Optimization}}}
Topology optimization enhances swarm robustness and longevity. Recent works employ evolutionary algorithms such as Gray Wolf and moth flame, and reinforcement learning for clustering and adaptive topology control \cite{alam_topology_2022}, \cite{feng_resilience_2022}, \cite{zhou_topology_2023}, \cite{fang_distributed_2024}. \cite{alam_topology_2022} and \cite{zhang_uav-enabled_2024} demonstrate improved stability and energy efficiency using multi-objective optimization to balance robustness and mission success. However, these approaches rarely integrate cross-layer dependencies or adversarial modeling, limiting their resilience against targeted attacks.

The rest of this paper is organized as follows: Section II details our proposed three-layer network model and adversarial model. Section III presents the SurBi-Ranking method, a GCN-based approach for assessing node and edge criticality. Section IV describes the multi-objective topology optimization using the NSGA-III algorithm. Section V conducts simulation experiments to evaluate the performance of the proposed framework. Section VI concludes our work.

\section{System Architecture}
This section establishes a comprehensive system model to address critical oversimplifications in existing swarm topological structures that neglect communication-layer vulnerabilities. We then propose a senior adversarial model that details criticality-aware and structured attack capabilities of the malicious attacker, so as to support our subsequent real-time criticality evaluation. Finally, to capture the evolving nature of extended operations, we present a multi-mission phase framework that integrates phase-specific networks with accelerated‑failure‑time factors, delivering a holistic, temporally aware assessment of node importance across all mission stages.

\subsection{Network Model}
\begin{figure}[h]
    \centering
    \includegraphics[width=\columnwidth]{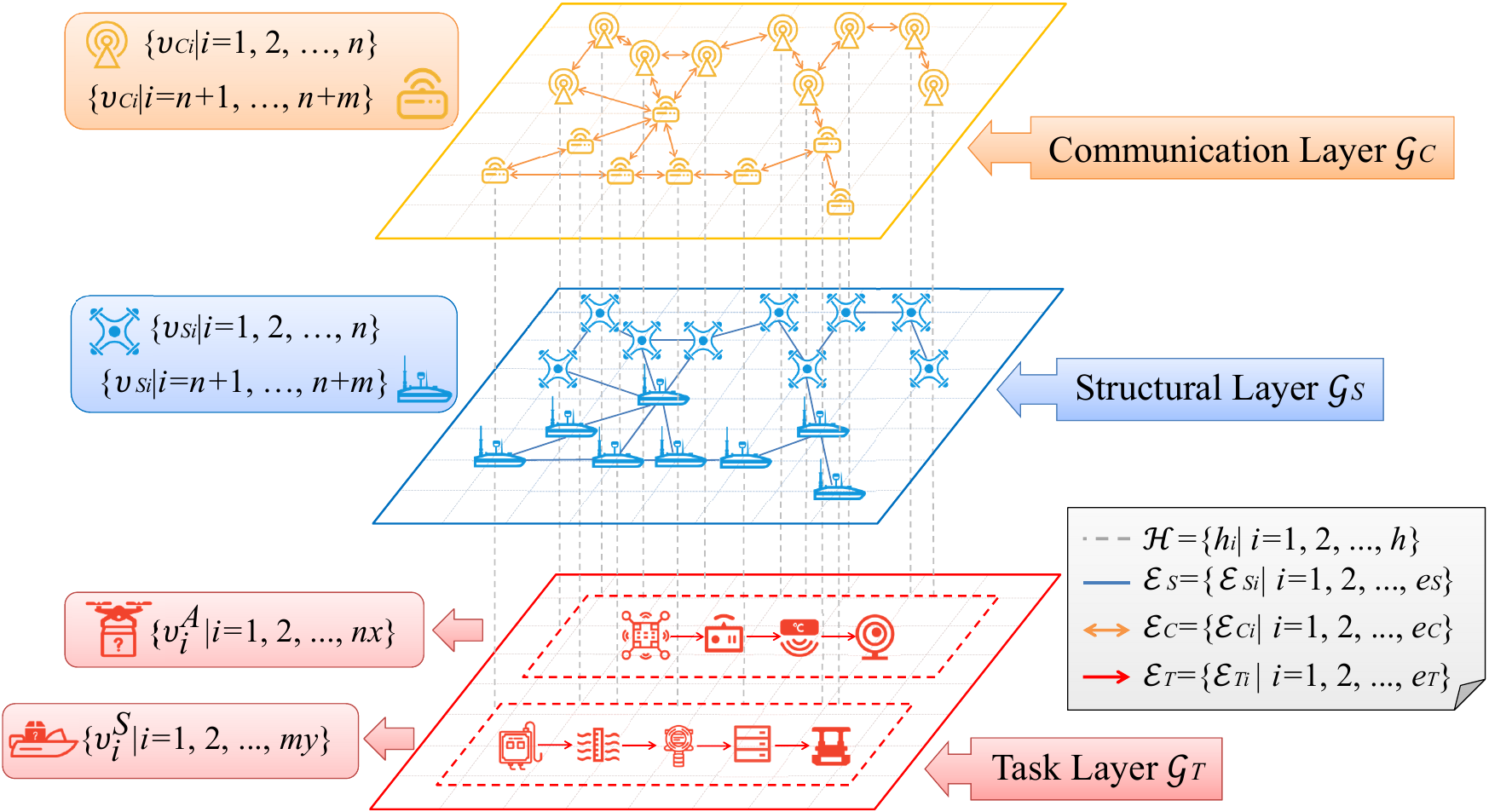}
    \caption{The architecture of our proposed three-layer network.}
    \label{fig2}
\end{figure}
As shown in Fig. \ref{fig2}, we construct a communication-structure-task coupled three-layer complex network $\mathcal{G}=\left(\mathcal{V}, \mathcal{E}\right)=\{\mathcal{G}_L|L=C,S,T\}$. The indices $C$, $S$ and $T$ in the node set $\mathcal{V}=\{v_{Li}| L=C, S, T; i=1, 2, ...v_L\}$ represent the communication modules, unmanned vehicles, and their payloads, respectively. We suppose that the cluster has $n$ UAVs and $m$ USVs, thus $\{v_{Si}|i=1,2,...,n\}$ and $\{v_{Si}|i=n+1,n+2,...,n+m\}$ represents the UAVs and USVs of the structure layer, respectively. There are totally different $z$ types of payloads, where each UAV has $x$ payloads, and each USV has $y$ payloads. The loads on different UAV/USVs are the same and are recorded as $\mathcal{V}_{T}=\{v_{i}^{V}| V=A,i=1,2,...,nx;V=S,i=1,2,...,my\}$, superscript $A$ representing the UAV and $S$ representing the USV. We assume that the communication module of each unmanned vehicle corresponds to a node on the communication layer $\mathcal{G}_C$, each unmanned vehicle corresponds to a node on the structure layer $\mathcal{G}_S$, and each payload corresponds to a node on the task layer $\mathcal{G}_T$, respectively. This three-layer complex network includes intra-layer edges $\mathcal{E}=\{e_{Li}| L=C,S,T;i=1,2,...,e_L\}$ and inter-layer edges $\mathcal{H}=\{h_i| i=1,2,...,h\}$.

\subsubsection{\textbf{Communication Layer}}
The communication layer serves as a dynamic mapping of the structural layer's unmanned vehicles, where nodes explicitly represent the onboard communication modules of these vehicles. Edges $\mathcal{E} \subseteq \mathcal{V} \times \mathcal{V}$ in this layer correspond to wireless links between these modules, with properties such as latency, bandwidth, and bit error rate directly reflecting real-world communication constraints. Crucially, each node in the communication layer is a functional abstraction of its structural counterpart such as the physical UAV/USV, thereby establishing a bidirectional dependency. The reliability of edges in this layer is governed by distance-dependent probabilistic failures, interference effects, and adversarial attack likelihoods. For instance, the link failure probability $P_{\text{fail}}(d)$ between two nodes modeled by an exponential failure probability model is denoted as
\begin{equation}\label{1}
P_{\text{fail}}(d)\ =\ 1-\mathrm{e}^{-(\frac{d}{d_0})^{n}},
\end{equation}
where the parameters are defined as follows:
\begin{itemize}
     \item $d$: Actual distance between nodes.
     \item $d_\mathrm{0}$: Characteristic distance.
     \item $n$: Cath loss exponent (free space $n=2$, complex environment $n=4\sim6$).
\end{itemize}

The communication success rate $\varOmega(i,j)$ between nodes $v_i$ and $v_j$ is then defined as the maximum reliability of all available paths:
\begin{equation}\label{2}
\varOmega(i,j) = \max_{p \in \mathcal{P}_{ij}} \prod_{e \in p} \underbrace{(1 - P_{\text{fail}}(d_e))}_{W(d_e)},
\end{equation}
where $\mathcal{P}_{ij}$ denotes path sets connecting $v_i$ and $v_j$, and $W(d_e) = 1 - P_{\text{fail}}(d_e)$ represents the edge weight derived from \eqref{1}. This formulation captures the optimal end-to-end reliability achievable in the network.

\subsubsection{\textbf{Structural Layer}}
The structural layer serves as the topological backbone of the unmanned vehicle swarm, explicitly encoding the physical connections and formation relationships among UAVs and USVs. Nodes in this layer represent individual vehicles, while edges correspond to formation-maintaining interactions, such as relative positioning constraints or collaborative maneuvers. This layer dynamically adapts to environmental disturbances including wind gusts, ocean currents and mechanical failures. Understanding the connectivity and robustness of this layer is crucial for assessing the overall resilience of the swarm, which can be quantified through key network metrics such as node degree.

The degree $k$ of a node is generated by another node on the edge connected to it, which may come from the layer where the node itself is located or from other layers. In the structure layer, UAVs have $1+x$ lines and USVs have $1+y$ lines. So there is the following expression $k_{Li}=k_{Li}(\mathcal{G}_{L})+k_{Li}^{\ell},L=C,S,T;i=1,2,...,v_{L}$, where  $k_{Li}(\mathcal{G}_{L})$ represents the number of edges from the layer where the other node is located, $k_{Li}^{\ell}$ represents  the number of edges from  the other node across layers, which is given by 
\begin{equation}
k_{Li}^{\ell}=
\begin{cases}
1,& \mathcal{V}_{Li}\in \mathcal{G}_{T} \\
1+x,& \mathcal{V}_{Li}\in \mathcal{G}_{C}\mid \mathcal{G}_{S}, i\leq n \\
1+y,& \mathcal{V}_{Li}\in \mathcal{G}_{C}\mid \mathcal{G}_{S}, n<i\leq m+n 
\end{cases}.
\end{equation}
The average degree and degree distribution can be derived, which is give by
\begin{equation}
\langle k\rangle=\frac{1}{v_C+v_S+v_T} \sum_{L=C, S, T} \left( \sum_{i=1}^{v_L} k_{Li}\right).
\end{equation}
The degree distribution function represents the proportion of nodes with a certain degree or more, or the probability $P_T(k)=\sum_{k^{\prime} \geq k} P\left(k^{\prime}\right)$. Obviously , the above degree, average degree, and degree distribution are all related to the task stage, and each task stage needs to be calculated based on the actual situation.

\subsubsection{\textbf{Task Layer}}
In practice, the workflow of the swarm is complex, which is called a multi-phase mission \cite{gu_group_2022}. 
\begin{figure}[h]
    \centering
    \includegraphics[width=0.95\columnwidth]{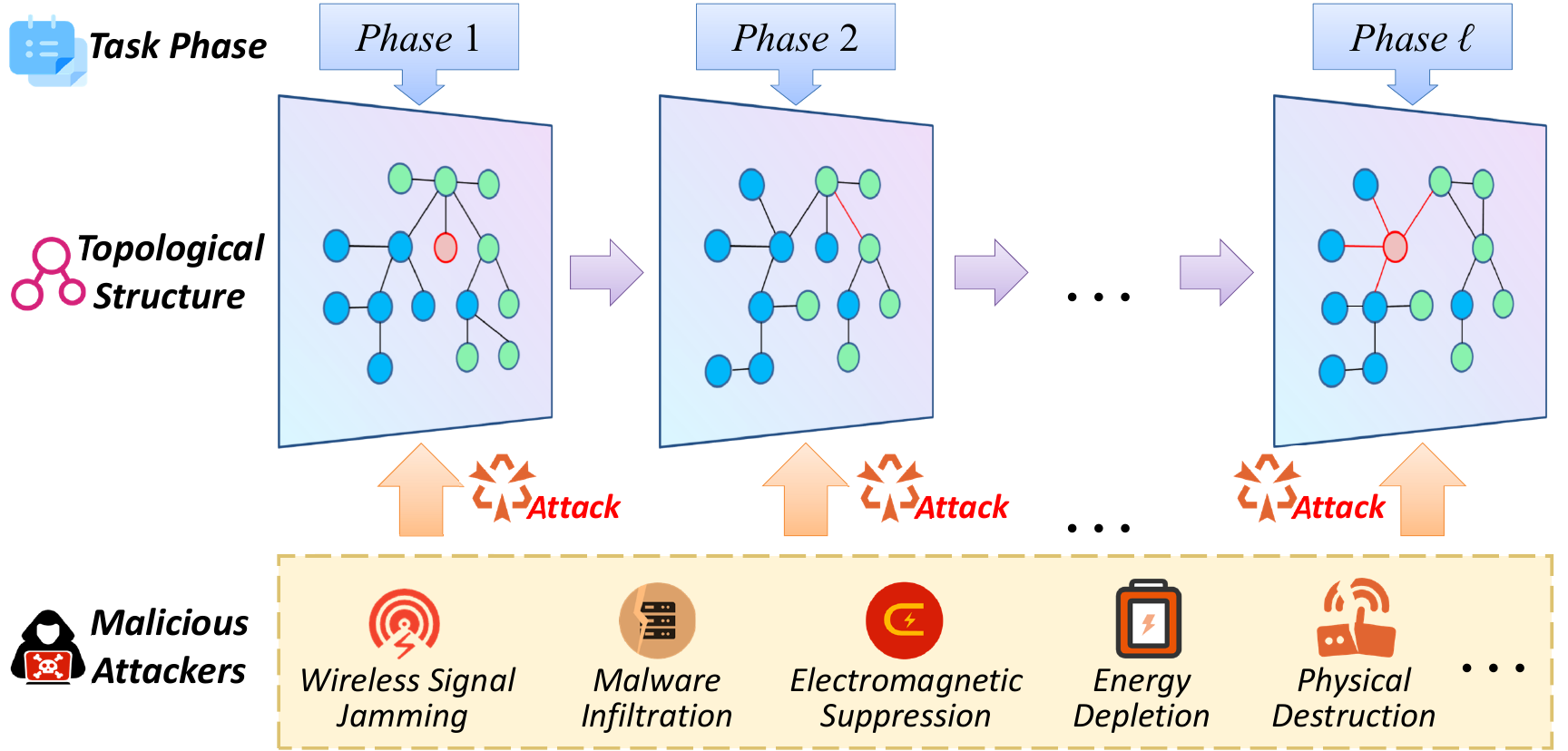}
    \caption{Multi-phase mission under malicious threats.}
    \label{fig3}
\end{figure}

As shown in Fig. \ref{fig3}, the underlying network topology $\mathcal{G} = (\mathcal{V}, \mathcal{E})$ remains invariant throughout all mission phases $Phase = [Phase\,1, Phase\,2, \ldots, Phase\,\ell]$. For each phase $j$, a designated \textit{active subgraph} $\mathcal{G}_j = (\mathcal{V}_j, \mathcal{E}_j)$ is activated, where $\mathcal{V}_j \subseteq \mathcal{V}$ denotes the subset of unmanned vehicles required for phase-specific tasks, and $\mathcal{E}_j = \{ e \in \mathcal{E} | e \text{ connects } u,v \in \mathcal{V}_j \}$ represents the operational communication links. This activation mechanism satisfies:
\begin{equation}
|\mathcal{V}_j| = m_j + n_j, \sum_{j=1}^\ell |\mathcal{V}_j| \geq |\mathcal{V}|, \mathcal{V}_i \cap \mathcal{V}_j \neq \emptyset \text{ for } i \neq j,
\end{equation}
where $m_j$ UAVs and $n_j$ USVs are required for phase $j$, allowing platform reuse across phases but requiring topological consistency throughout the mission lifecycle.

For each active subgraph $\mathcal{G}_j$, phase-specific vulnerability is quantified via percolation critical probability. For random ER networks with Poisson-distributed degrees $P(k) = e^{-\langle k \rangle} \langle k \rangle^k / k!$, the critical probability simplifies to:
\begin{equation}
P_{Tj} = \frac{1}{k_{0j} - 1}, \quad k_{0j} = \frac{\langle k^2 \rangle_j}{\langle k \rangle_j},
\end{equation}
demonstrating uniform vulnerability where random and targeted attacks yield comparable impacts. Here $\langle k \rangle_j$ and $\langle k^2 \rangle_j$ are computed solely over $\mathcal{G}_j$.  Conversely, scale-free networks with power-law degree distributions $P(k) = ck^{-\gamma}$ exhibit topology-dependent vulnerability:
\begin{equation}\label{7}
k_0 \to \left| \frac{2-\gamma}{3-\gamma} \right| \times 
\begin{cases} 
k_{\min}, & \gamma > 3 \\
k_{\min}^{\gamma-2}k_{\max}^{3-\gamma}, & 2 < \gamma < 3 \\
k_{\max}, & 1 < \gamma < 2 
\end{cases}
\end{equation}

The phase fragility metric $P_j$ combines structural robustness with payload reliability:
\begin{equation}
P_j = \underbrace{\left[1 - \text{Normalization}(P_{Tj})\right]}_{\text{structural}} \times \underbrace{\prod_{i \in \mathcal{V}_j} R_i(t_j)}_{\text{functional}}.
\end{equation}
Payload reliability $R_i(t_j)$ follows the Accelerated Failure Time Model (AFTM) \cite{chen_mixture_2018}:
\begin{equation}
R_i(t_j) = \exp\left(-\delta_i \sum_{p=1}^{j} \xi_{ip}^{j} T_p\right)
\end{equation}
where $\delta_i$ is the base failure rate, $T_p$ is phase duration, and $\xi_{ip}^{j}$ are mission-stress acceleration factors. Higher $P_j$ indicates greater phase survivability within the static topology constraints.

Global mission success requires maintaining inter-phase connectivity through bridge nodes $\mathcal{V}_B = \bigcap_{j=1}^\ell \mathcal{V}_j$:
\begin{equation}
C_{\text{global}} = - \exp\left(-\sum_{j=1}^{\ell} \beta_j \cdot \lambda_2(\mathcal{G}_j) \cdot \mathbb{I}_{\text{conn}}(\mathcal{G}_j)\right),
\end{equation}
where $\lambda_2$ is algebraic connectivity, $\mathbb{I}_{\text{conn}}$ is the connectivity indicator (1 if $\mathcal{G}_j$ connected, 0 otherwise), and $\beta_j$ weights phase criticality. The mission success probability aggregates phase outcomes:
\begin{equation}
P_{\text{task}} = \prod_{j=1}^{\ell} P_j \cdot \exp\left(-\eta \sum_{u \in \mathcal{V}_B} \deg(u)\right).
\end{equation}
The exponential penalty term captures the cascading impact of bridge node failures, with $\deg(u)$ counting phases where $u$ participates.

\subsection{Adversarial Model}
\begin{figure}[h]
    \centering
    \includegraphics[width=0.95\columnwidth]{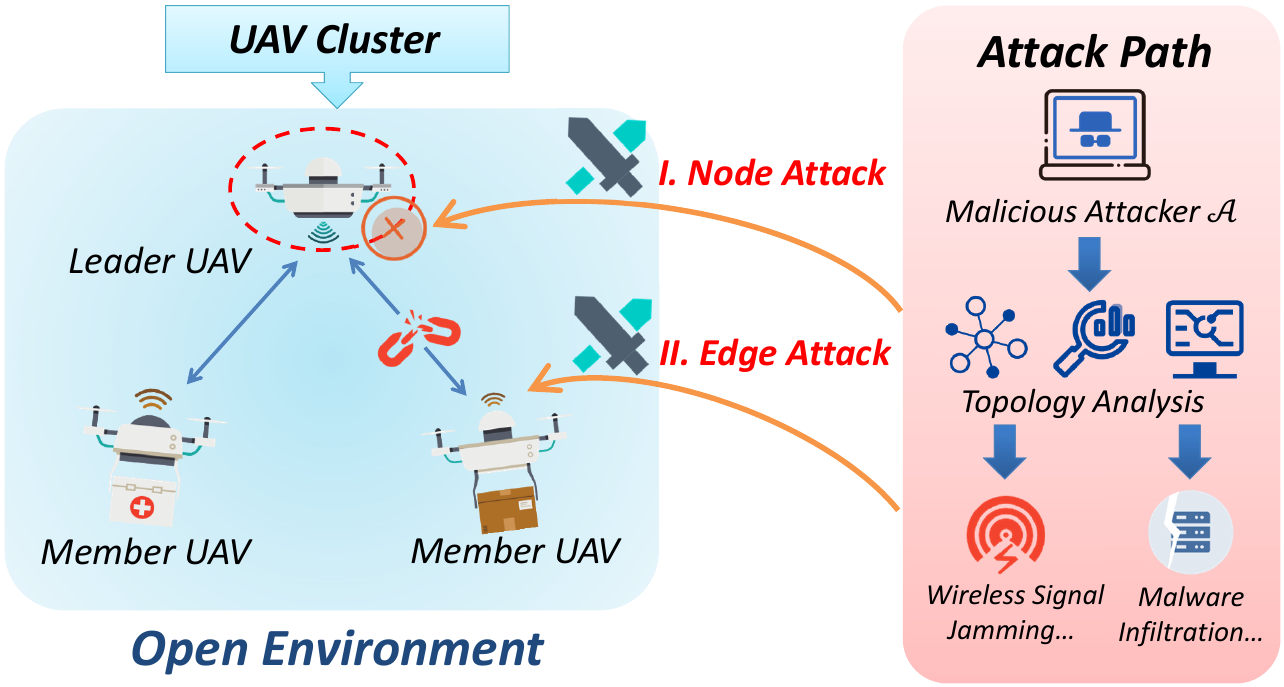}
    \caption{Attack pattern diagram of adversary.}
    \label{fig4}
\end{figure}
\begin{table*}[ht]
      \centering
      \caption{Malicious Attacker’s Capacities}
      \begin{center}
      \begin{tabular}{|c|l|l|}
        \hline
        \textbf{${CAP}_*$} & \textbf{Capability Type }&\textbf{Description of Adversary’s Capacities} \\
        \hline \hline
        ${CAP}_1$ & Situational Awareness & $\mathcal{A}$ can analyze the marine-aerial swarm topology $\mathcal{G}$ at any phase $j$.\\ \hline
        ${CAP}_2$ & Criticality Inference  & $\mathcal{A}$ can identify and prioritize high-value nodes or edges based on real-time network structure.\\ \hline
        ${CAP}_3$ & Node Disruption  & $\mathcal{A}$ is able to physically destroy, deplete energy, or inject malware into selected unmanned vehicles.\\ \hline
        ${CAP}_4$ & Edge Disruption  & $\mathcal{A}$ has the ability to sever communication links via jamming, side-channel attack, or electromagnetic interference.\\ \hline
        ${CAP}_5$ & Cross-Layer Strikes & $\mathcal{A}$ has the capability to induce cascading failures on structurally correlated components across layers. \\ \hline
      \end{tabular}
      
      \label{tab1}
      \end{center}
\end{table*}

In this section, We propose a criticality-aware and time-adaptive adversarial model featuring a powerful attacker $\mathcal{A}$ with real-time situational awareness (${CAP}_1$), as formalized in Table \ref{tab1}. Leveraging dynamic analysis of the marine-aerial swarm network’s topology and mission states (${CAP}_2$), $\mathcal{A}$ executes temporally coordinated attacks (${CAP}_5$), systematically eliminating nodes (${CAP}_3$) and edges (${CAP}_4$) to degrade connectivity and mission capacity \cite{liu_towards_2022}, \cite{gong_resilient_2025}.

\subsubsection{\textbf{Edge-Based Attack}} 
The attacker $\mathcal{A}$ can employ techniques including side-channel attack, wireless signal jamming, and electromagnetic suppression to dynamically sever communication links between specific unmanned vehicles, which results in the removal of edges $\mathcal{E}_{a} \subseteq \mathcal{E}_C$ from the communication layer $\mathcal{G}_C$:
\begin{equation}
\mathcal{G}_C \rightarrow \mathcal{G}_C' = \mathcal{G}_C \setminus \mathcal{E}_{a},
\end{equation}
where edge removal cascades to node failures in the structural layer when connectivity is fully lost.

\subsubsection{\textbf{Node-Based Attack}} 
By involving physical destruction attack or energy depletion attack to completely incapacitate specific unmanned vehicles, the attacker $\mathcal{A}$ targets critical nodes, such as leader nodes, in the marine-aerial swarm network, leveraging topology information to prioritize high-value targets and trigger cascading failures across layers \cite{feng_study_2022}. Disabling an unmanned vehicle carrying critical payloads disrupts dependent subtask chains, compromising multi-phase mission continuity, with the removal of the targeted node $v_a$ from the structural layer $\mathcal{G}_S$ propagating to the communication $\mathcal{G}_C$ and task layers $\mathcal{G}_T$ as follows: 
\begin{align}
& \quad \mathcal{G}'_S = \mathcal{G}_S - \{v_a\}, \\
& \quad \mathcal{G}'_C = \mathcal{G}_C - \{\phi(v_a)\} - \{e \in \mathcal{E}_C \mid \phi(v_a) \in e\}, \\
& \quad \mathcal{G}'_T = \mathcal{G}_T - \{u \in \mathcal{V}_T \mid \operatorname{host}(u) = v_a\},
\end{align}
where $\phi: \mathcal{V}_S \rightarrow \mathcal{V}_C$ is the bijective mapping from structural nodes to communication nodes, and $\operatorname{host}: \mathcal{V}_T \rightarrow \mathcal{V}_S$ assigns task nodes to their host vehicles.

This cross-layer propagation model captures the impact of node attacks on network connectivity and mission integrity.

\section{Criticality Evaluation}
In this section, we introduce a novel method for evaluating node importance. We begin by integrating complex network theory with graph convolutional networks to construct a powerful node feature–extraction model. Then, we develop a comprehensive evaluation framework to quantify node significance. This comprehensive evaluation method, which combines a node’s Birnbaum importance with the surrounding node importance, is termed Surrounding-Birnbaum Importance Ranking (SurBi-Ranking). Finally, we apply this approach to the multi-mission phase model described above and derive the importance ranking of nodes within the resulting multi-phase complex network.
\subsection{Erection of Graph Convolution Network}
\begin{figure*}[t]
            \centering
            \includegraphics[width=\textwidth]{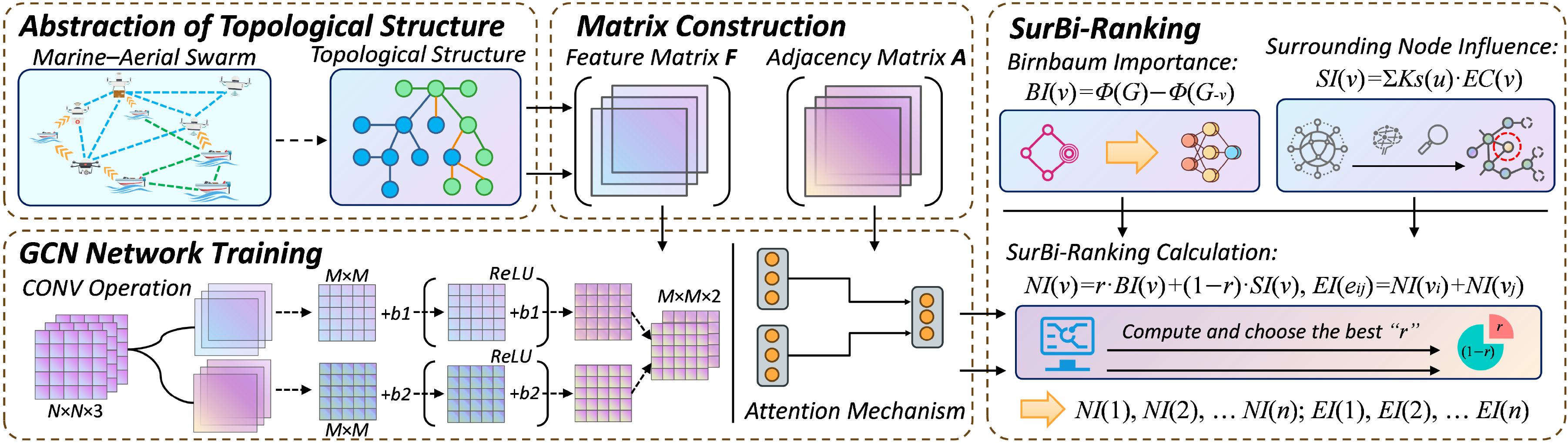}
            \caption{Architecture of our proposed \textit{SurBi-Ranking} method. The system constructs feature and adjacency matrices from the swarm topology, processes them through a GCN, and combines Birnbaum importance and surrounding node influence to compute real-time node criticality scores.}
            \label{fig5}
\end{figure*}

This section addresses the construction of evaluation models for key nodes in complex network models. The nodal characteristic matrix $\textbf{\textit{F}}$ and the adjacency matrix $\textbf{\textit{b}}$ are used as inputs to GCN. The output of the model is the insertion of nodes. Finally, the evaluation index of key nodes is constructed, and the nodes' importance is sorted by their importance scores. The flowchart of the GCN is shown in Fig. \ref{fig5}.
\subsubsection{Feature matrix} 
The feature matrix $\textbf{\textit{F}}$ consists of the topology feature. Using the centricity index to construct the topology feature. The five centricity indices are degree centricity (DC) \cite{jia_improvement_2019}, betweenness centrality (BC) \cite{veremyev_finding_2017}, eigenvector centricity (EC) \cite{bonacich_eigenvector-like_2001}, closeness centrality (CC) \cite{hu_closeness_2019} and K-shell value (KS) \cite{hu_k-shell_2015}. Therefore, the feature matrix $\textbf{\textit{F}}$ can be represented as
\begin{equation}\textbf{\textit{F}}=
\begin{bmatrix}
DC_1 & BC_1 & EC_1 & CC_1 & KS_1\\
DC_2 & BC_2 & EC_2 & CC_2 & KS_2\\
\vdots & \vdots & \vdots & \vdots & \vdots  \\
DC_i & BC_i & EC_i & CC_i & KS_i
\end{bmatrix}.\end{equation}
To make different features be the same order of magnitude, the features need to be normalized. The feature $val_i$ of the node $v_i$ is normalized, which is represented as 
\begin{equation}val_i=\frac{rank_i}{N},\end{equation}
where rank is the rankings of node $v_i$ scores according to indexes $DC$, $BC$, $EC$, $CC$ and $KS$, and $N$ is the total number of nodes.
\subsubsection{Adjacency matrix}
In a complex network, the adjacency matrix $\textbf{\textit{A}}$ is calculated by
\begin{equation}
A_{ij}=
\begin{cases}
1, & \text{If the nodes } i \text{ and } j \text{ bordered junction;} \\
0, & \text{Else.}
\end{cases}
\end{equation}
\subsubsection{The formula of GCN}
The core of the GCN aims to extend the convolution operation to graph structured data and updates the node representation by converging the characteristics of the node and its neighbors. The following is the most classical forward propagation GCN, which is written as
\begin{equation}
H^{(l+1)}=\sigma\left(\hat{D}^{-1/2}\hat{A}\hat{D}^{-1/2}H^{(l)}W^{(l)}\right),
\end{equation}
 where $\hat{A}$ is the adjacency matrix containing a self-ring, i.e. $\widehat{A}=A+I$, the symbol $\hat{D}$ is the degree matrix, $H^{(l)}$ is the nodal characteristic matrix of layer $l$, $W^{(l)}$ is the trainable weight matrix, and \(\sigma(\cdot)\) is an activation function, e.g.,\ ReLU function. The GCN has a total of $L$ layers.

\subsubsection{Model training}
To ensure efficient convergence of the GCN, we employ neighbor sampling combined with mini-batch stochastic gradient descent (SGD) with momentum to optimize the model parameters. Let the input graph be \(\mathcal{G}=(\mathcal{V},\mathcal{E})\), where each node \(v_i\in \mathcal{V}\) has an initial feature vector \(x_i\) and label \(y_i\in\{1,\dots,C\}\).

For each training node \(v_i\), at the \(l\)‑th layer, we randomly sample up to   \(K_l\) of its first‑order neighbors
\begin{equation}
\mathcal{N}^l(v_i)=\mathrm{Sample}\bigl(\mathcal{N}(v_i),\,K_l\bigr),  
\end{equation}
where \(\mathcal{N}(v_i)\) denotes the full neighbor set of \(v_i\).

After \(L\) layers, we apply a final linear projection and softmax to obtain class probabilities \(\hat p_{i,c}\). We optimize the cross-entropy loss over the training node set \(\mathcal{V}_{\mathrm{train}}\) with the loss function
\begin{equation}
\mathcal{L}= -\frac{1}{|\mathcal{V}_{\mathrm{train}}|}\sum_{v_i\in\mathcal{V}_{\mathrm{train}}}\sum_{c=1}^C \mathbb{I}[y_i = c]\,\ln \hat p_{i,c}\,.
\end{equation}

\subsection{Node Importance}
This section focuses on the node importance assessment, including Birnbaum importance and Surrounding node influence. In turn, we get the importance score for each node.
\subsubsection{Birnbaum Importance}
Birnbaum importance measure was originally derived from the field of reliability engineering to assess the importance of individual components in a system to overall system reliability. We apply this concept to complex networks, expanding it to assess how critical network nodes are to the overall connectivity of the network. The Birnbaum importance measure reflects how important a node in a network is to the overall network. The specific expression can be written as
$$BI(v)=\varPhi(\mathcal{G})-\varPhi(\mathcal{G}_{-v}),$$
where $\varPhi(\mathcal{G})$ is the global natural connectivity of the network $\mathcal{G}$, and $\varPhi(\mathcal{G}_{-v})$ is the natural connectivity after removing node $v$. $\varPhi(\mathcal{G})$ is the natural connectivity of the network $\mathcal{G}$, and its expression is represented as
\begin{equation}\varPhi(\mathcal{G})=\ln\left(\frac{1}{N}\sum_{k=1}^Ne^{\lambda_k}\right),\end{equation}
where $\{\lambda_k\}_{k=1}^N$ is all eigenvalues of the adjacency matrix of $\mathcal{G}$.
\subsubsection{Surrounding Node Influence}
In complex networks, the importance of nodes around a node is also a key measure of the node's importance, which represents a local feature. Here, we not only consider the $K_s$ values of a single node, but often several different nodes have close $K_s$ values when there are many nodes. At this time we consider the neighboring nodes of each node, and the sum of $K_s$ values of all the neighboring nodes of a node is about, so the node is also a relatively important node. The sum of the $K_s$ value is denoted as
\begin{equation} KS(v) =\sum_{u\in\varGamma\left(v\right)}K_s\left(u\right),\end{equation}
where $\varGamma\left(v\right) $ is the neighborhood of node $v$, referring to the neighbor node of $v$.
Eigenvector centricity is also an important measure of the importance of surrounding nodes. Thus, we define a new composite indicator $SI(v)$. Its expression is represented as
\begin{equation}SI(v)=KS(v)\cdot EC(v).\end{equation}
\subsubsection{Surrounding-Birnbaum Importance
Ranking}
Here, we propose the SurBi-ranking node criticality evaluation method, which we define as a new metric that fuses global and local perspectives to assess node importance in real time.
The SurBi-ranking is calculated from Birnbaum importance and surrounding node influence, which is given by
\begin{equation}NI(v)=r\cdot BI(v)+(1-r)\cdot SI(v),\end{equation}
where $r$ is the weight assigned to Birnbaum importance. Uniquely, we propose an adversarial evaluation-based approach to determine its optimal value. Specifically, we simulate targeted attacks on the marine-aerial swarm using node rankings generated under different $r$ values, and identify the value of $r$ that leads to the fastest system degradation or infection propagation. 

Specifically, we define the importance of an edge as the sum of the importance scores of the two nodes it connects. This formulation captures the idea that an edge is more critical if it links two highly important nodes. Formally, edge importance is computed as
\begin{equation}
EI(e_{ij}) = NI(v_i) + NI(v_j),
\end{equation}
where $EI(e_{ij})$ denotes the importance of edge $e_{ij}$, and $NI(v_i), NI(v_j)$ are the SurBi-Ranking scores of the connected nodes $v_i$ and $v_j$, respectively. This approach ensures that the SurBi-ranking metric captures the most critical nodes and edges from an attacker’s perspective, thereby maximizing its effectiveness in real-time resilience assessment.

\subsection{Multi Stage Evaluation}

At each mission stage, we use the SurBi-Ranking method to evaluate the nodes in the mission layer and get their real-time importance scores. In this case, we define the global importance of each node
\begin{equation}\phi(v)=-e^{-\sum_{j=1}^{l}\beta_{j}\cdot NI_{j}(v)},\end{equation}
where $NI_{j}(v)$ is the complex network importance of the node $v$ in phase $j$. The symbol $\beta_{j}$ is the acceleration coefficient for phase $j$, which can be estimated by fitting historical failure/outage time data or based on experience settings. The symbol $\phi(v)$ is the cumulative acceleration factor of the node $v$, which can reflect the importance of the node $v$ after the end of all mission phases.

\section{Dynamic Topology Optimization}
\subsection{Initial Static Topology Optimization}
The global objective function can be diveded into three sub-objectives, which are
\begin{enumerate}
    \item \textbf{Global Connectivity($f_{1}$):} Measured via global algebraic connectivity $\lambda_2(\mathcal{G})$ of the Laplacian matrix $\textbf{\textit{L}}=\textbf{\textit{D}}-\textbf{\textit{A}}$:
     \begin{equation}
        f_{1} = \lambda_2(\mathcal{G}).
    \end{equation}
    Considering that the network needs to be fully connected, we exclude the case of $\lambda_2(\mathcal{G})=0$.
    \item \textbf{Global Communication Success Rate ($f_{2}$)}: 
    \begin{equation}
        f_{2} = \frac{2\sum_{i\neq j}\varOmega(v_i,v_j)}{N(N-1)}.
    \end{equation}
    \item \textbf{Global Vulnerability ($f_{3}$):} Complement of the Task Success Rate $P_{\text{task}}$:
    \begin{equation}
         f_{3} =1 - \prod_{j=1}^{\ell}P_{j}\cdot \exp\left(-\eta \sum_{u \in \mathcal{V}_B} \deg(u)\right).
    \end{equation}
\end{enumerate}

As per the objective functions, we use NSGA-III to optimize the swarm topology
\begin{equation}
    \min_{\mathcal{G}} \left\{ 1-f_{1}, 1-f_{2}, f_{3}\right\}.
\end{equation}
The algorithm generates a 3D Pareto front $\mathcal{P}_N$ for each edge count $|\mathcal{E}| = N$. The optimal topologies are selected from $\mathcal{P}_N$ by using TOPSIS for decision making, which can be given by
\begin{enumerate}
    \item Define weight vectors $\mathbf{w}_i = [w_{\text{1}}^i, w_{\text{2}}^i, w_{\text{3}}^i]$ with $\sum w_{\bullet}^i = 1$. For each $\mathbf{w}_i$, compute the TOPSIS score for all solutions in $\mathcal{P}_N$.
    \item Select the topology $\mathcal{G}_i^*$ with minimal TOPSIS score.
    \item Perform \textit{targeted attacks} on $\mathcal{G}_i^*$: Remove nodes in descending order of $NI(v)$ and record the connectivity drop curve, and the solution with slowest decay determines the optimal $\mathbf{w}_i^*$ and $\mathcal{G}^*$.
\end{enumerate}

\subsection{Dynamic Topology Adjustment under Attack}
\begin{algorithm}
\caption{Dynamic Topology Reconfiguration}
\label{alg1}
\begin{algorithmic}[1]
\REQUIRE

       $\mathcal{G}_0$: Compromised graph \\
       \qquad \,\: $j_a$: Attack phase \\
       \qquad \,\; $N$: Target edge count

\ENSURE Optimal topology $\mathcal{G}^{*}$

\STATE Generate candidate graphs $\{\mathcal{G}_k\}$ with edge count $\approx N$
\STATE Compute node importance $NI(v)$ by SurBi-Ranking

\FOR{each $\mathcal{G}_k$}
    \STATE Calculate:
    \STATE \quad Global connectivity $f_1$
    \STATE \quad Global communication success rate $f_2$
    \STATE \quad Subsequent Vulnerability $f_3'$
    \STATE \quad Reconfiguration cost $f_4$
\ENDFOR

\STATE Find Pareto solutions $\mathcal{P}$ by NSGA-III:\\
    \qquad$\min_{\mathcal{G}'} \{1-f_1, 1-f_2, f_3', f_4\}$

\FOR{each solution in $\mathcal{P}$}
    \STATE Simulate attacks by removing nodes with $NI \downarrow$ 
    \STATE Record connectivity decay rate
\ENDFOR

\STATE Select solution with slowest decay as $\mathcal{G}^{*}$

\RETURN $\mathcal{G}^{*}$
\end{algorithmic}
\end{algorithm}
As provided in Algorithm \ref{alg1}, we extend the method above to scenarios where nodes $\mathcal{V}_a$ (with $\mathcal{E}_{va}$ incident edges) and edges $\mathcal{E}_a$ are attacked at phase $j_a$:
\begin{enumerate}
    \item Let $\mathcal{G}_0 = \mathcal{G} \setminus (\mathcal{V}_a \cup \mathcal{E}_{va} \cup \mathcal{E}_a)$ be the compromised graph.
    \item Generate new graphs $\{\mathcal{G}_k'\}$ with $|\mathcal{E}'| \approx N$.
    \item Change the $3^{rd}$ objective function \textbf{Subsequent Vulnerability ($f_{3}'$)}:
    \begin{equation}
         f_{3}' =1 - \prod_{j=j_a}^{\ell}P_{j}\cdot\exp\left(-\eta \sum_{u \in \bigcap_{j=j_a}^\ell \mathcal{V}_j} \deg(u)\right).
    \end{equation}
    \item Introduce the $4^{th}$ objective \textbf{Reconfiguration Cost ($f_{4}$)}:
    \begin{equation}
        f_{\text{4}} =|\mathcal{E}_{k}'|+|\mathcal{E}_0|-2|\mathcal{E}_{k}' \cap \mathcal{E}_0|.
    \end{equation}
    \item Solve 4D optimization via NSGA-III:
\begin{equation}
    \min_{\mathcal{G}_k'} \left\{ 1-f_{1}, 1-f_{2}, f_{3}, f_{4}\right\}.
\end{equation}
    \item Apply TOPSIS with updated weights to identify optimal post-attack topology $\mathcal{G}^*$.
\end{enumerate}

\section{Experimental and simulation verification}
In this section, we conduct simulation experiments to implement and evaluate the previously benchmarked methods.

\subsection{DataSets and Experimental Setup}
We employ three datasets in our experimental framework:
\subsubsection{\textbf{Planar Layer Network (PLN)}} 
\textit{DataSet I} simulates solutions for the $r$ value using a scale-free Planar Layer Network in~\eqref{7} with 1000 nodes following a power law distribution, selected for its connectivity properties that enable attack simulations. 
\subsubsection{\textbf{Multi-Phase Mission Dataset}} Identical to \textit{DataSet I}, \textit{DataSet II} models a marine-aerial swarm network executing a five-phase mission with a fixed set of 1000 nodes. Each mission phase features distinct graph topologies due to dynamically reconfigured edge connections under similar topological constraints, only the optimal connected subgraph that satisfies phase-specific requirements for UAV/USV quantities and configuration participates in task execution, while non-participating nodes remain inactive.
\subsubsection{\textbf{3D Contested Environment Dataset}} \textit{DataSet III} facilitates topological optimization of heterogeneous swarms in contested 3D environments, comprising 50 nodes (30 UAVs and 20 USVs) distributed within a 1000 m $\times$ 1000 m $\times$ 1000 m volume under layered heterogeneity: USVs occupy the sea surface ($z=0$) while UAVs operate at $z \in [50,1000]~\text{m}$. Network configurations incorporate variable edge densities while ensuring globally connected structures. Inter-node distances are computed using 3D Euclidean metrics. The swarm executes a multi-phase mission with stage-specific operational requirements, enabling robustness-reliability-mission success trade-off analyses under adversarial conditions.

\subsection{SurBi-Ranking Parameter Optimization}
\begin{figure}[h]
\includegraphics[width=\columnwidth]{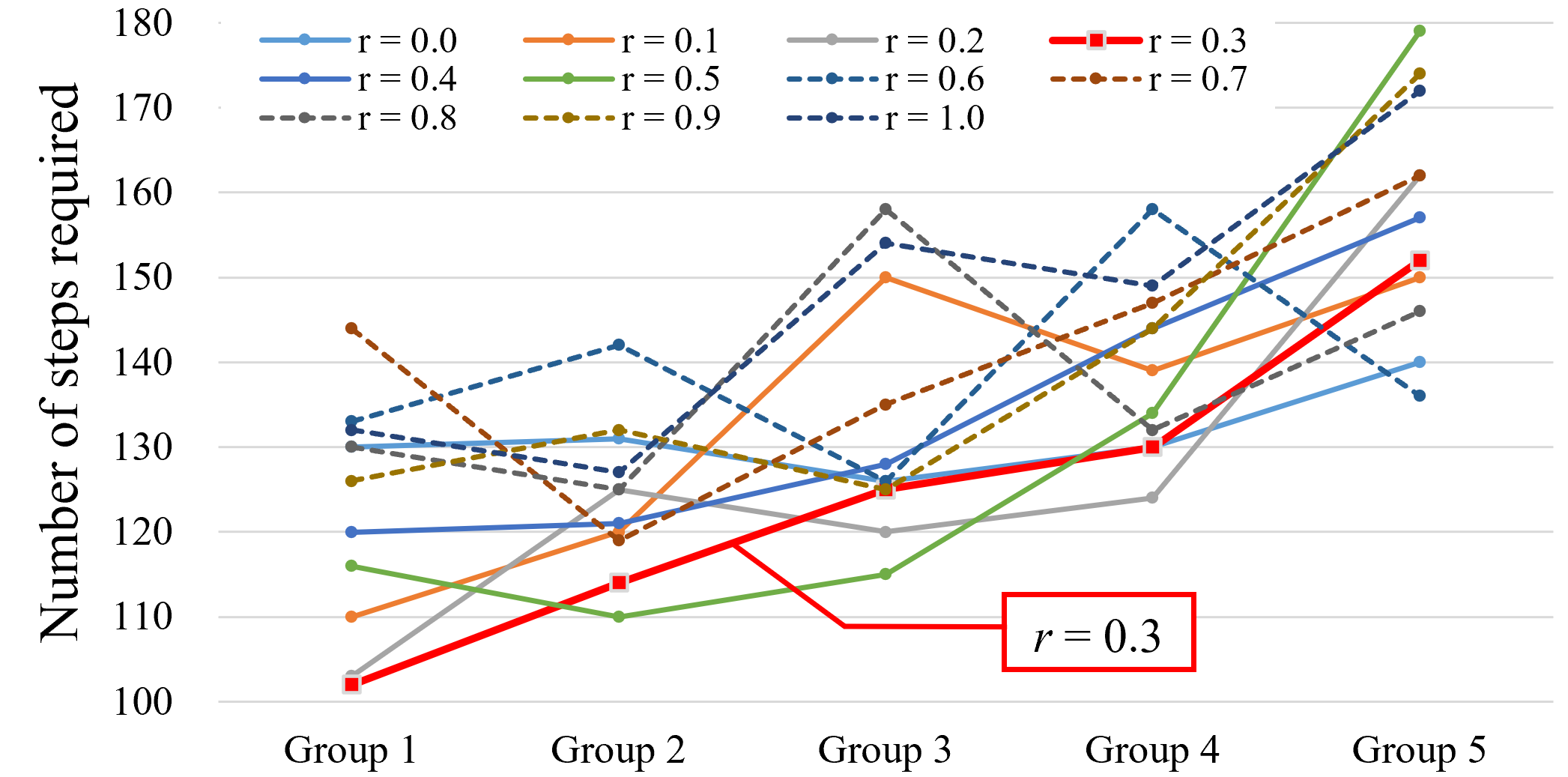}
\caption{SIR test with different $r$ values.}
\label{fig6}
\end{figure}
Different $r$ values correspond to different importance rankings. Under a certain $r$ value, the top 250 SurBi-Ranked nodes are divided into 5 groups in \textit{DataSet I}. We use the SIR model \cite{Haenggi2015The}, which is commonly used for infectious disease transmission, to test the results of the significance assessment. At some specific $r$ value, we initially infected each group separately and measured the time taken when the network had the most infected nodes. The results are shown in Fig. \ref{fig6}.

In this experiment, we want to find the best $r$ value corresponding to the curve should meet the following conditions: (1) From Group 1 to Group 5, the corresponding time should continue to rise, which illustrates the $r$ value under the group is reasonable, (2) The overall time for all groups, especially Group 1, should be as small as possible. As shown in the figure, the above requirements are satisfied when $r=0.3$, which is the optimal $r$ value for this swarm.

Critically, the optimal $r$ value is topology-dependent, and distinct swarms require case-specific calibration for their unique $r$ values.
\subsection{Superiority Evaluatuin of SurBi-Ranking}
\begin{figure*}[th]
    \centering
    \begin{minipage}[t]{0.46\textwidth}
        \centering
        \begin{subfigure}[b]{0.48\textwidth}
            \includegraphics[width=\textwidth]{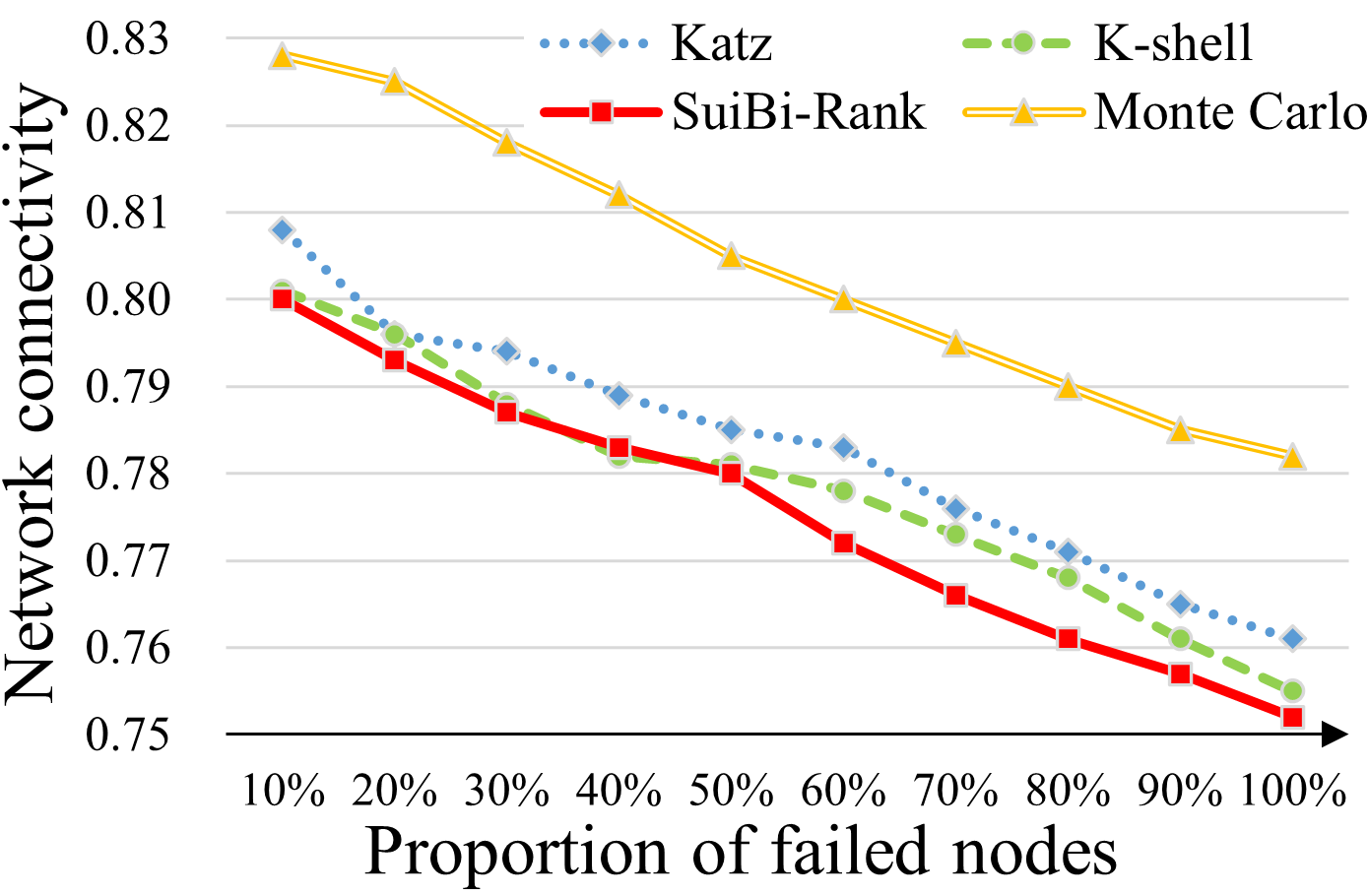}
            \caption{}
            \label{fig7a}
        \end{subfigure}
        \hfill
        \begin{subfigure}[b]{0.48\textwidth}
            \includegraphics[width=\textwidth]{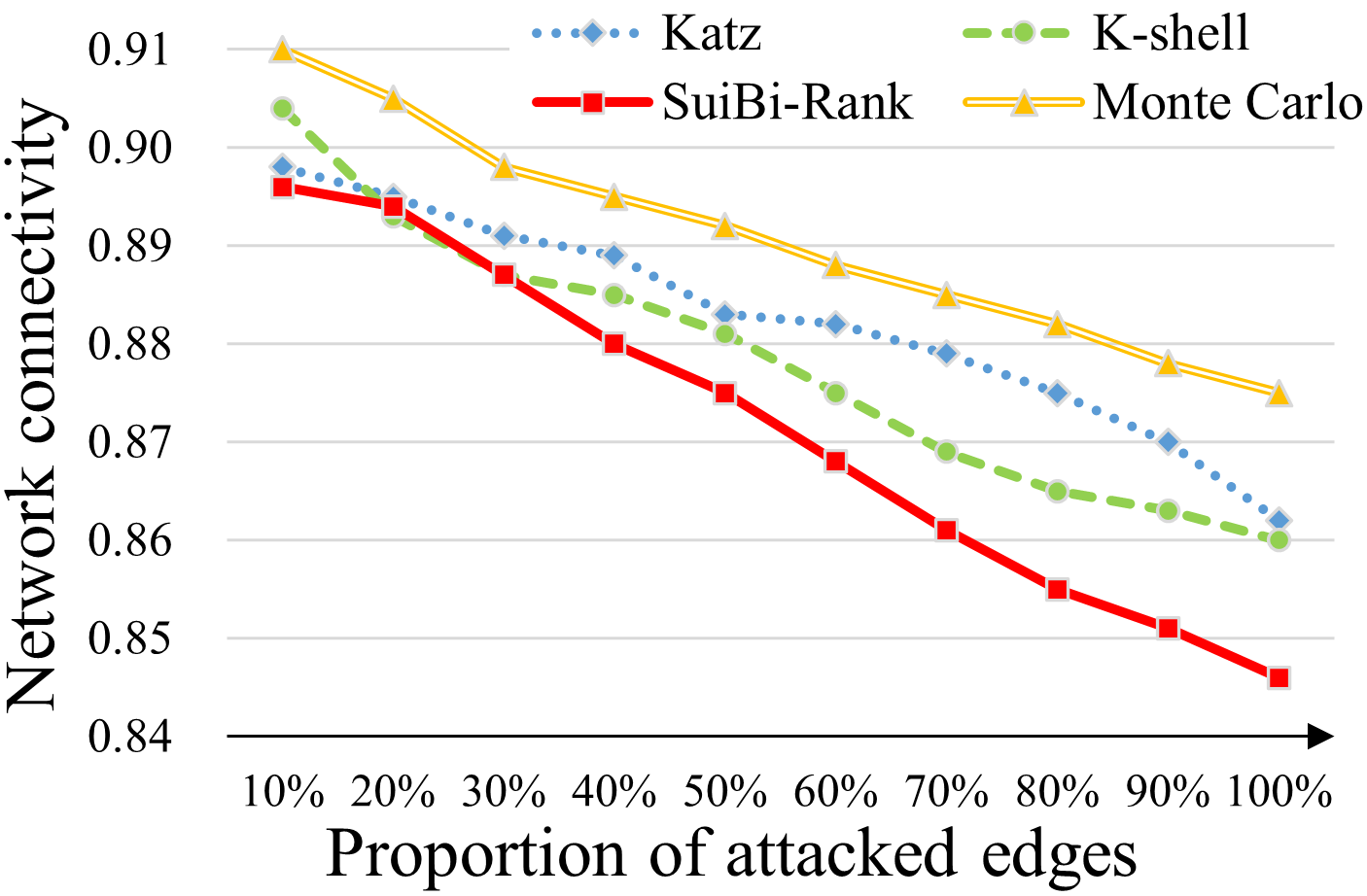}
            \caption{}
            \label{fig7b}
        \end{subfigure}
        \caption{PLN Networks' connectivity degradation under two types of attacks: (a) node attacks and (b) edge attacks.}
        \label{fig7}
    \end{minipage}
    \hfill
    \begin{minipage}[t]{0.24\textwidth}
        \centering
        \includegraphics[width=\textwidth]{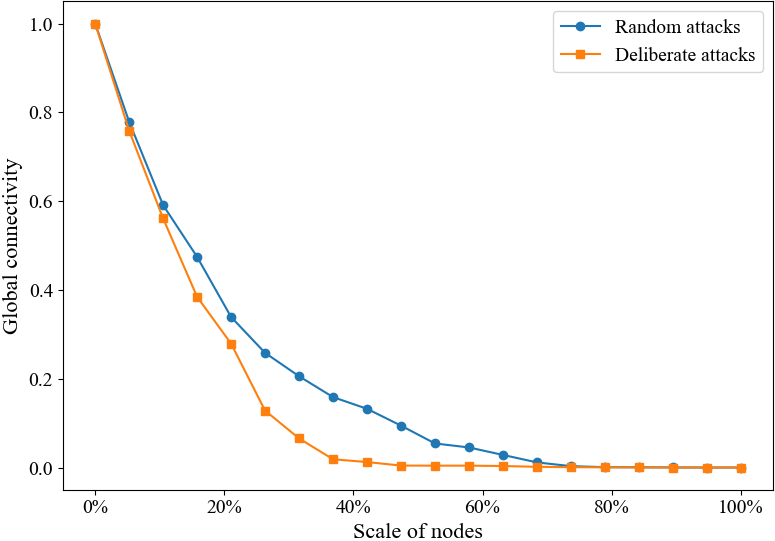}
        \caption{Global connectivity under random/deliberate attacks.}
        \label{fig8}
    \end{minipage}
    \hfill
    \begin{minipage}[t]{0.26\textwidth}
        \centering
        \includegraphics[width=\columnwidth]{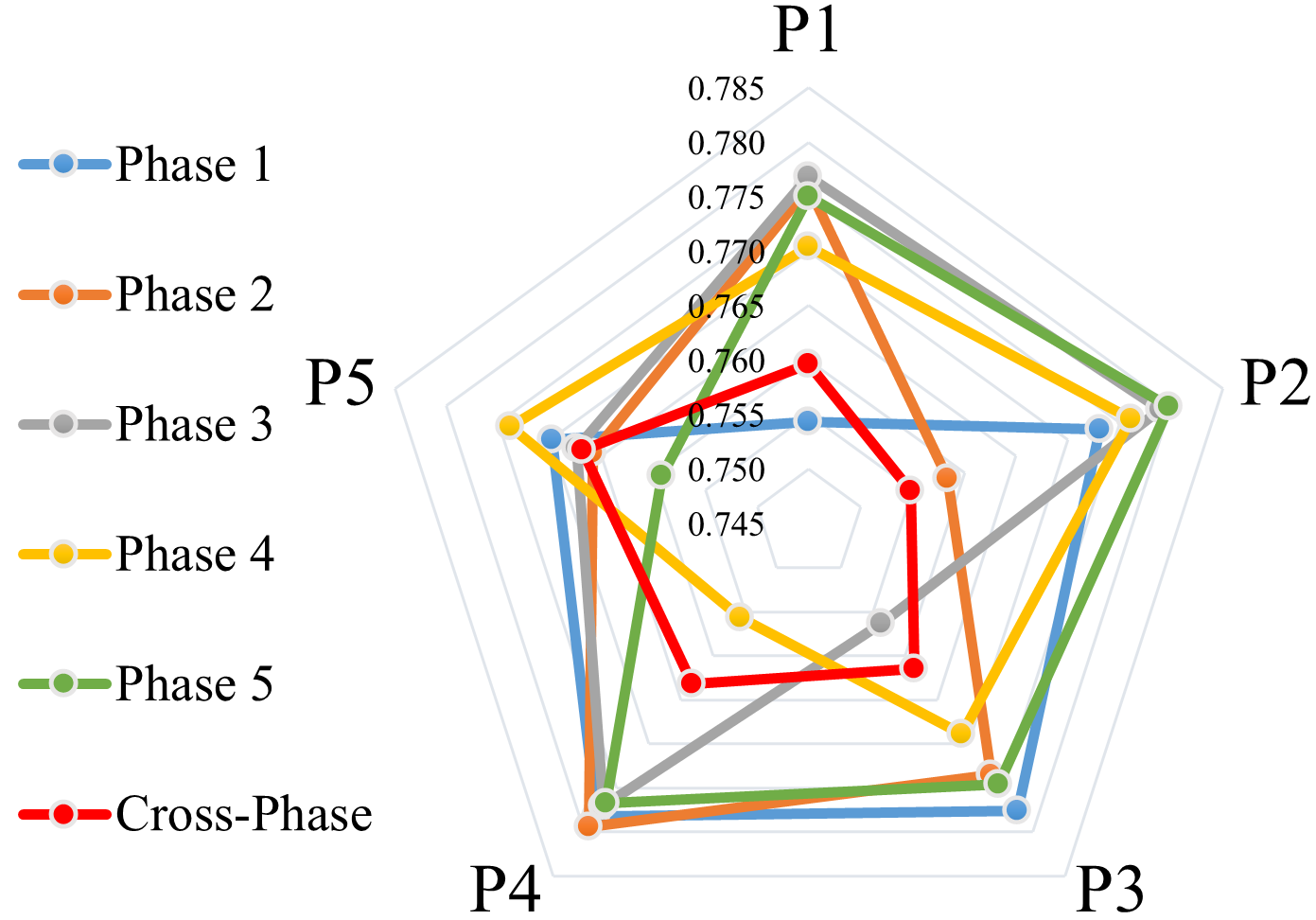}
    \caption{The multi-phase mission connectivity radar chart.}
    \label{fig9}
    \end{minipage}
\end{figure*}

To validate the accuracy of the key components identified by SurBi-ranking, we compared its performance against two classical centrality measures: K-shell and Katz, and the Monte Carlo method is used to simulate random attacks as a control group. Under both node-attack and edge-attack scenarios, we first computed node-importance scores using each method and then successively removed nodes in descending order of importance. As the number of removed nodes increased, we monitored how the network’s connectivity evolved. 

As shown in Fig. \ref{fig7}, regardless of whether the network was subjected to node attacks or edge attacks, the decline in network connectivity was consistently steeper when nodes were removed according to SurBi-ranking than when removed according to either K-shell or Katz centrality. This indicates that SurBi-ranking more accurately identifies the most critical nodes, thus conferring a clear advantage over the other two methods. We attribute this superiority to the fact that neither the K-shell nor the Katz index offers a fully comprehensive metric: SurBi-ranking, by contrast, captures not only the intrinsic importance of each node but also the collective influence exerted by its neighbors. Furthermore, we observe that targeted attacks based on importance ranking cause a more rapid degradation of network connectivity than random attacks.

\subsection{Criticality Analysis of Multi-Phase Mission}
We assess global node importance across the five mission-phase graphs in \textit{DataSet II}, obtaining importance rankings for all nodes. Fig. \ref{fig8} compares global connectivity degradation under two attack scenarios: Monte Carlo random node failures versus deliberate attacks based on SurBi-Ranking. The deliberate attack curve exhibits significantly steeper connectivity decline, validating our global criticality assessment efficacy.

Next, after analyzing the criticality of the entire mission, the adversary $\mathcal{A}$ disables the top 10\% critical nodes in each phase along with the top 10\% critical nodes across the entire multi-phase mission. The radar chart in Fig.~\ref{fig9} reveals that removing phase-specific critical nodes causes sharp connectivity degradation in their respective phases: Phase 1-5 drop to around $0.75$, with less cross-phase impact. Conversely, attacking globally critical nodes induces a uniform connectivity decline across all phases, demonstrating their systemic influence on mission continuity. This aligns with percolation theory: phase-specific nodes cause localized disruptions while cross-phase critical nodes pose system-wide risks, necessitating phase-aware criticality assessment for resilient swarm planning.

\subsection{Static Topology Optimization}
\begin{figure*}[htbp]
\centering
\begin{minipage}[t]{0.235\textwidth}
\centering
\includegraphics[width=\columnwidth]{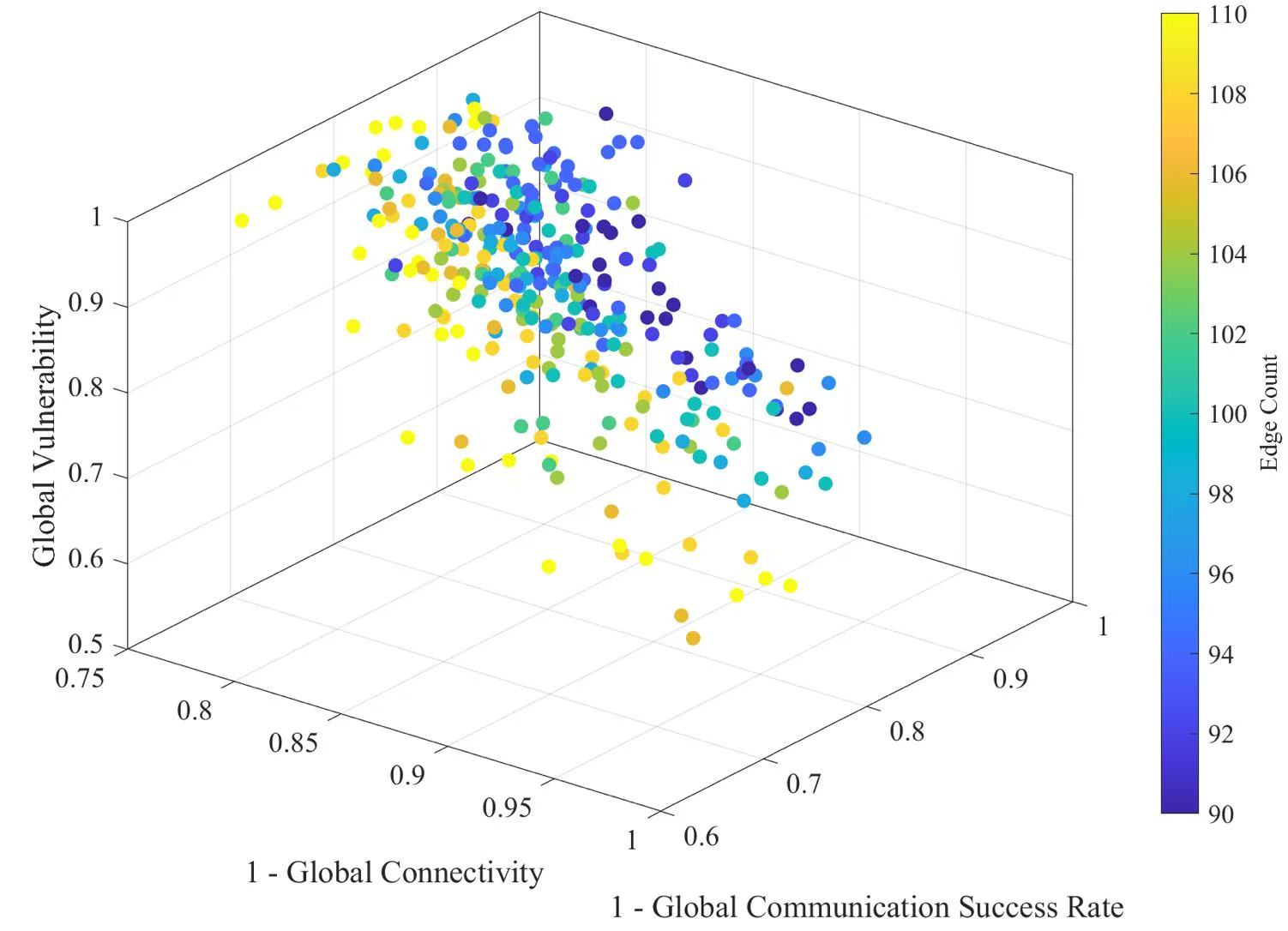}
\caption{3D Pareto frontier under NSGA-III optimisation.}
\label{fig10}
\end{minipage}
\hfill
\begin{minipage}[t]{0.26\textwidth}
\centering
\includegraphics[width=\columnwidth]{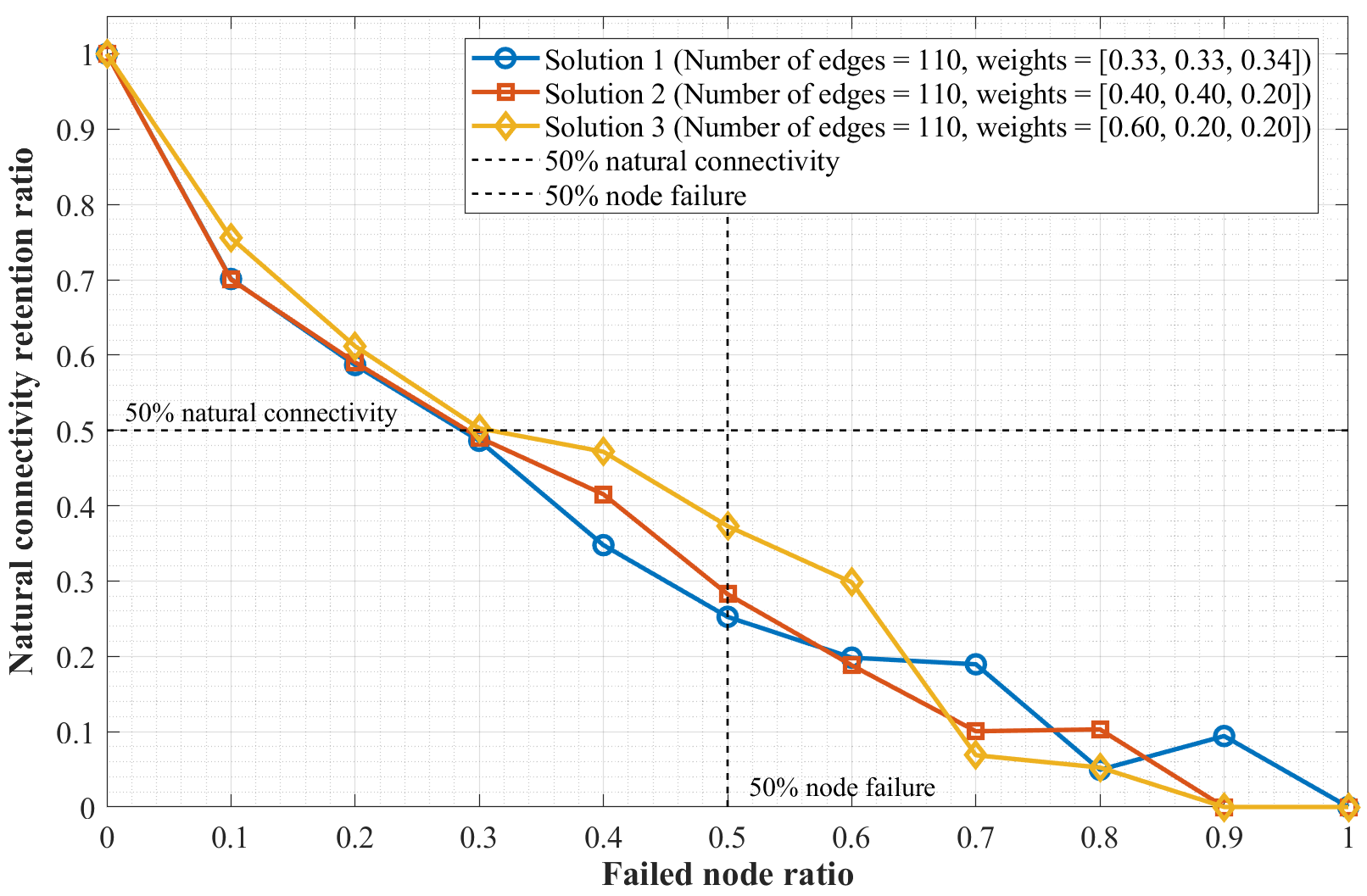}
\caption{TOPSIS for static optimization attack resilience.}
\label{fig11}
\end{minipage}
\hfill
\begin{minipage}[t]{0.26\textwidth}
\centering
\includegraphics[width=\columnwidth]{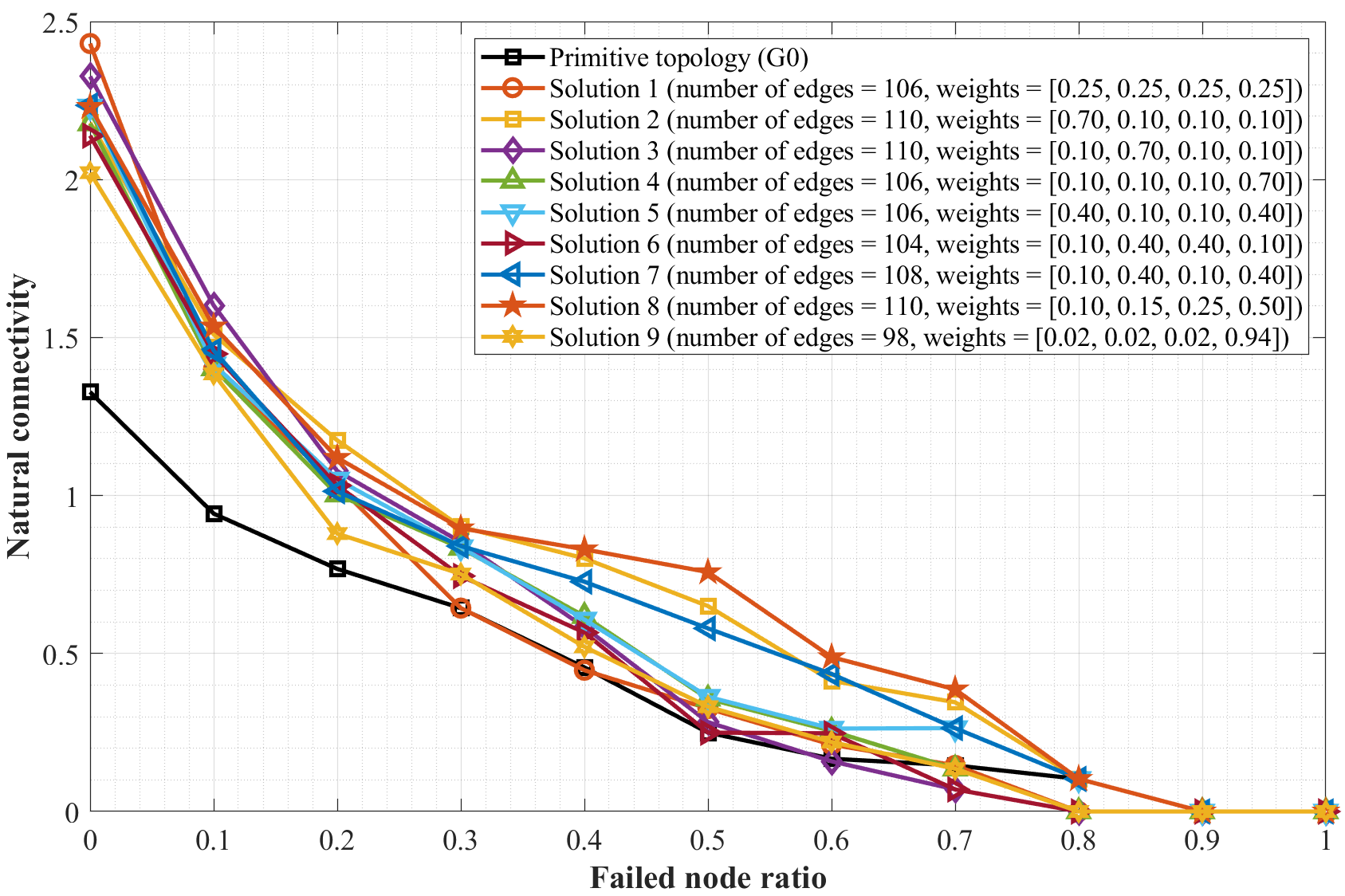}
\caption{TOPSIS for dynamic reconfiguration resilience.}
\label{fig12}
\end{minipage}
\hfill
\begin{minipage}[t]{0.22\textwidth}
\centering
\includegraphics[width=\columnwidth]{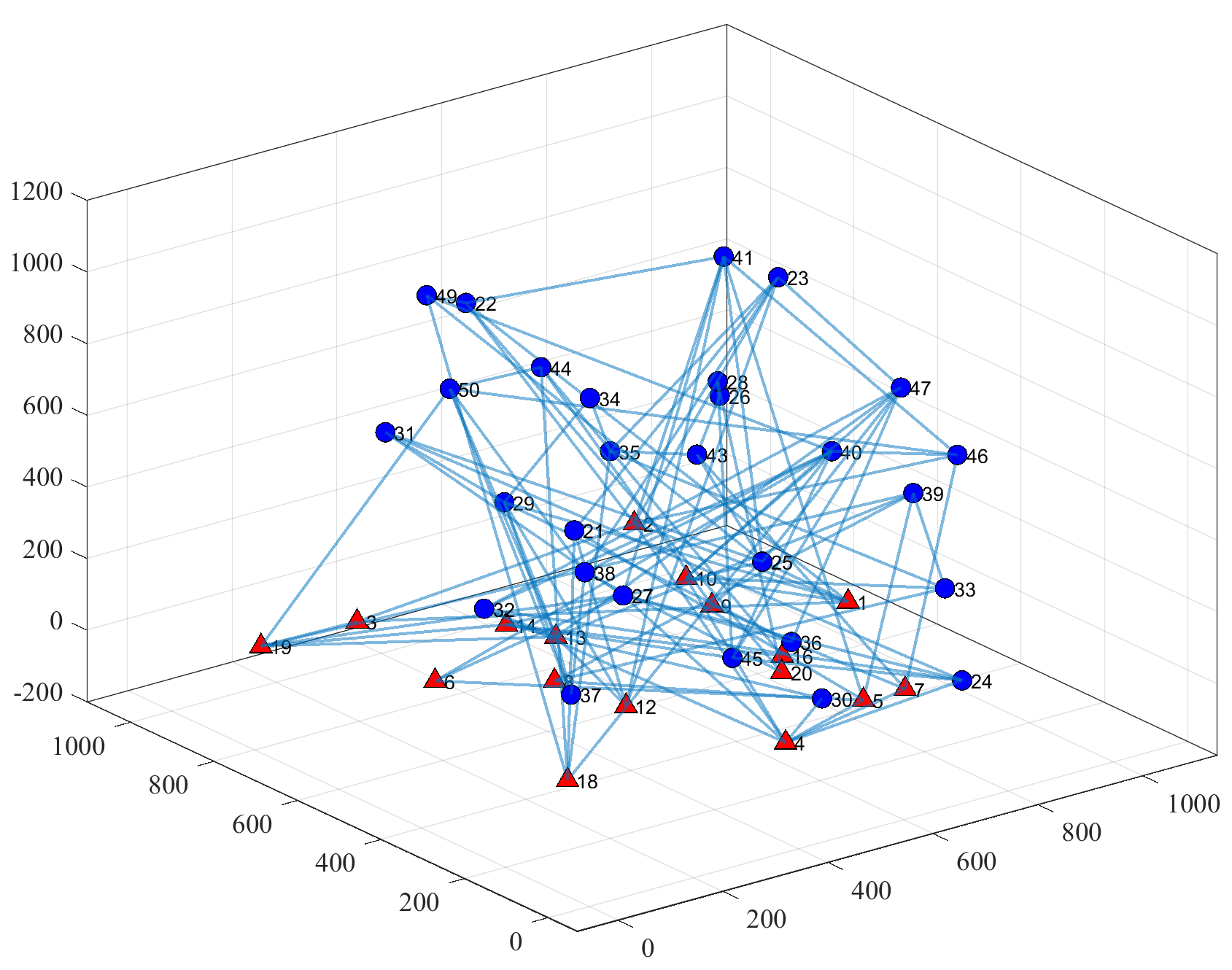}
\caption{Optimal 3D structure graph of the swarm.}
\label{fig13}
\end{minipage}
\end{figure*}
We employ the NSGA-III algorithm to minimize three objectives $\min_{\mathcal{G}} \left\{ 1-f_{1}, 1-f_{2}, f_{3}\right\}$ to obtain the 3D Pareto frontier, and the results are shown in Fig. \ref{fig10}. Then, we employ the TOPSIS method to determine optimal network topologies under varying weight assignments for multiple optimization objectives. For each optimal solution corresponding to distinct weight distributions, we conduct simulated attacks by sequentially disabling nodes in descending order of their SurBi-Ranking importance. During this process, a fixed proportion of nodes is eliminated at each attack step. The solution exhibiting the slowest degradation rate of natural connectivity is identified, along with its associated weight assignment configuration. The results are shown in Fig. \ref{fig11}.

The solution analysis identifies Solution 3 as optimal, with the objective weight distribution:
\[
\mathbf{w}_3 = \left[ w_{1}^3, w_{2}^3, w_{3}^3 \right] = \left[ 0.60, 0.20, 0.20 \right]
\]

\subsection{Dynamic Topology Optimization}
Given a specific topology graph $\mathcal{G}$ of an unmanned cluster in \textit{DataSet III}, the graph fails due to an attack on a set of nodes $\mathcal{V}_a$ and edges $\mathcal{E}_a$. Based on the compromised graph $\mathcal{G}_0$, we perform the dynamic optimization process by minimize four objectives $\min_{\mathcal{G}_k'} \left\{ 1-f_{1}, 1-f_{2}, f_{3}, f_{4}\right\}$. For all solutions in the 4D Pareto frontier $\mathcal{P}_N$, we apply TOPIS with varying weight distributions $\mathbf{w}_i$ over optimization metrics, yielding optimal topology sets $\{\mathcal{G}_k^*\}$.

For the optimal solution under each different weight distribution at this point, we apply the simulated attack by attacking nodes in descending order of SurBi-Ranking importance, causing a fixed proportion of nodes to fail each time. This yields the solution with the slowest decline in natural connectivity and the weight distribution under that solution, which are the optimal solution and optimal weight distribution, respectively. The results are shown in Fig. \ref{fig12}.

As can be seen from Fig. \ref{fig12}, all solutions in the optimal solution set of the reconstructed topology significantly enhance the natural connectivity beyond $\mathcal{G}_0$. The optimal solution in the solution set $\{\mathcal{G}_k^*\}$ represents the optimal topology dynamic adjustment strategy after attack. Based on the analysis of the natural connectivity in the solution and its downward trend after the attack, solution~8 is determined to be the optimal solution, with the following target weight distribution
\[
\mathbf{w}_8 = \left[ w_1^8, w_2^8, w_3^8, w_4^8 \right] = \left[ 0.10, 0.15, 0.25, 0.50 \right],
\]
and the optimal 3D structure graph of the unmanned aerial-marine swarm after attack is shown in Fig. \ref{fig13}.

In contrast to other solutions, the optimal solution~8 balances the competing objectives by prioritizing mission-critical connectivity ($w^{8}_3 = 0.25$) while minimizing retasking overhead ($w^{8}_4 = 0.50$), reducing the natural connectivity degradation by around 30\%, compared to the Pareto average. Moreover, solution 8 maintains a natural connectivity of 0.76 at 50\% critical-node failure, demonstrating higher resilience under extreme attacks. These results align with our objective of strengthening aerial-marine swarm networks against targeted disruptions, offering a robust solution for multi-phase missions under adversarial conditions.

\section{Conclusion}
This paper presents a robust framework to enhance the resilience of heterogeneous marine-aerial swarm networks against adversarial threats. We developed a three-layer network model capturing structural, communication, and task dependencies, paired with a sophisticated adversarial model simulating targeted attacks. The proposed SurBi-Ranking method, leveraging graph convolutional networks, accurately identifies critical nodes and edges, while NSGA-III-based topology optimization balances robustness, reliability, and mission success. Experimental results demonstrate superior performance over traditional methods, ensuring sustained connectivity and mission effectiveness.

\end{document}